\documentclass[twocolumn,superscriptaddress]{revtex4}
\usepackage{mathrsfs,braket}
\usepackage{amssymb, amsbsy, amsmath, latexsym, dsfont, array, layout,graphicx,mathrsfs,color,ulem,bm}
\usepackage{upgreek}
\usepackage{ulem}

\usepackage[colorlinks,citecolor=blue]{hyperref}

\allowdisplaybreaks

\begin{document}

\title{{\color{black}Emergent $\mathcal{PT}$-symmetry breaking of collective modes with topological critical phenomena}}

\author{Jian-Song Pan}
\email{panjsong@gmail.com}
\affiliation{Department of Physics, National University of Singapore, Singapore 117542, Singapore}

\author{Wei Yi}
\email{wyiz@ustc.edu.cn}
\affiliation{CAS Key Laboratory of Quantum Information, University of Science and Technology of China, Hefei 230026, China}
\affiliation{CAS Center For Excellence in Quantum Information and Quantum Physics, Hefei 230026, China}

\author{Jiangbin Gong}
\email{phygj@nus.edu.sg}
\affiliation{Department of Physics, National University of Singapore, Singapore 117542, Singapore}

\maketitle

\begin{center}{\bf Abstract}\end{center}
%\begin{abstract}
%{\bf {\color{black}
The spontaneous breaking of parity-time ($\mathcal{PT}$) symmetry yields rich critical behavior in non-Hermitian systems, and has stimulated much interest, albeit most previous studies were performed within the single-particle or mean-field framework.
%exploring the interplay between $\mathcal{PT}$ symmetry and quantum fluctuations in a many-body setting is a burgeoning frontier.
Here, by studying the collective excitations of a Fermi superfluid with $\mathcal{PT}$-symmetric spin-orbit coupling, we uncover an emergent $\mathcal{PT}$-symmetry breaking in the Anderson-Bogoliubov (AB) collective modes,
%whose spectra undergo a transition from being completely real to completely imaginary,
even as the superfluid ground state retains an unbroken $\mathcal{PT}$ symmetry. {The critical point of the transition is marked by a non-analytic kink in the speed of sound, which derives from the coalescence and annihilation of the AB mode and its hole partner, reminiscent of the particle-antiparticle annihilation.
The system consequently becomes immune to low-frequency external perturbations at the critical point, a phenomenon
{associated with} the spectral topology of the complex quasiparticle dispersion. This critical phenomenon offers a fascinating route toward perturbation-free quantum states.
%}}
%}
%\end{abstract}

%skyrmions
\begin{center}{\bf Introduction}\end{center}

The eigenspectrum of a parity-time ($\mathcal{PT}$)-symmetric Hamiltonian is either completely real, or formed by complex conjugate pairs, depending on the symmetry of its eigenstates~\cite{bender1998real,el2018non,bender2018pt}.
By tuning system parameters, the $\mathcal{PT}$ symmetry of eigenstates can be spontaneously broken across critical (or exceptional) points~\cite{moiseyev2011non}, where coalescing eigenstates and eigenenergies give rise to intriguing critical phenomena.
$\mathcal{PT}$-symmetry breaking and the critical phenomena thereof have been extensively studied in the past decades, over a plethora of physical systems ranging from photonics~\cite{feng2017non,pile2017gaining, limonov2017fano, horiuchi2017marriage,xiao2019observation,klauck2019observation,szameit2011p,regensburger2012parity,zhen2015spawning,weimann2017topologically, kim2016direct,cerjan2016exceptional,poshakinskiy2016multiple}, acoustics and phononics~\cite{cummer2016controlling}, to single spins~\cite{wu2019observation, liu2020dynamically}, quantum gases~\cite{li2019observation} and superconducting wires~\cite{rubinstein2007bifurcation,chtchelkatchev2012stimulation}.
Most of these prior studies relied on single-particle or mean-field descriptions.
Moving forward,
the interplay of $\mathcal{PT}$-symmetry breaking and many-body correlations, which lies at the cutting edge of the current research,
is expected to yield rich and exotic critical behavior~\cite{ashida2017parity,hanai2019non,zhou2019interaction,hanai2020critical,pan2020interaction,fruchart2020phase}.

{In previous studies, critical behavior at an exceptional point largely emerge in some exotic dynamics
 such as the enhanced spectral response~\cite{wiersig2014enhancing,wiersig2020review} and the robust energy transfer~\cite{sid2017robust, huber2019active,partanen2019exceptional}. In a many-body setting, experimentally relevant dynamic processes are generally dominated by low-energy excitations. Investigating the low-energy collective excitations in a $\mathcal{PT}$-symmetric many-body system is therefore a crucial first step toward a deeper understanding of the many-body criticality therein.}

In this work, we theoretically demonstrate an emergent $\mathcal{PT}$-symmetry breaking in the collective modes of a Fermi superfluid, and investigate in detail the rich many-body critical phenomena therein.
Specifically, we study the pairing superfluid and collective excitations of a two-component Fermi gas under a non-Hermitian, $\mathcal{PT}$-symmetric spin-orbit coupling (SOC). Characterized by a non-Hermitian extension of the Bardeen-Cooper-Schrieffer (BCS) theory, the ground state of the system is a $\mathcal{PT}$-symmetry-preserving superfluid with real energy. Intriguingly though, the Bogoliubov quasiparticle excitations above the BCS state feature complex dispersions, forming closed spectral loops on the complex plane.
The ground state can therefore be regarded as a point-gap topological superfluid, insofar as it possesses both the pairing order and a spectral winding topology  ~\cite{gong2018topological,kawabata2019symmetry,bergholtz2019exceptional,okuma2020topological,zhang2020correspondence}
regarding its quasiparticle excitations.

{Remarkably, the Anderson-Bogoliubov collective modes of the superfluid undergo a $\mathcal{PT}$-symmetry transition as the SOC strength is tuned, while the superfluid ground state remains $\mathcal{PT}$-symmetry unbroken. In particular, a critical SOC strength exists, separating $\mathcal{PT}$-symmetry unbroken and broken phases of the Anderson-Bogoliubov (AB) modes that have purely real or imaginary spectra, respectively. At this emergent $\mathcal{PT}$ transition, the AB mode and its hole partner coalesce and annihilate each other, leading to the complete disappearance of low-frequency excitations, as the speed of sound vanishes in a kink at the transition.
This is in sharp contrast to the case with a single-particle $\mathcal{PT}$-symmetric system, where eigenmodes merely merge at the critical point.
Such a many-body critical behavior is associated with the point-gap topology of the quasiparticle dispersion, which
suggests a topologically robust critical state that is immune to low-frequency perturbations.}

\begin{center}{\bf Results}\end{center}

{\bf Non-Hermitian Fermi gases with $\mathcal{PT}$-symmetric SOC.} We consider a two-component, attractively interacting Fermi gas in three dimensions. The Fermi gas is loaded in an optical lattice and subject to a one-dimensional, imaginary SOC, with the Hamiltonian
\begin{equation}\label{eq:origin_H}
\begin{split}
H=&-\sum_{\boldsymbol{k}}C_{\boldsymbol{k}}^{\dagger}(\sum_{\zeta=x,y,z}t_{\text{s},\zeta}\cos k_{\zeta}+it_{\text{so}}\sigma_{z}\sin k_{x})C_{\boldsymbol{k}}\\
&-\frac{U}{V}\sum_{\boldsymbol{k},\boldsymbol{k}^{'},\boldsymbol{q}}c_{\boldsymbol{q}+\boldsymbol{k}\uparrow}^{\dagger}c_{\boldsymbol{q}-\boldsymbol{k}\downarrow}^{\dagger}c_{-\boldsymbol{q}-\boldsymbol{k}^{'}\downarrow}c_{-\boldsymbol{q}+\boldsymbol{k}^{'}\uparrow},
\end{split}
\end{equation}
where $C_{\boldsymbol{k}}=(\begin{array}{cc} c_{\boldsymbol{k}\uparrow} & c_{\boldsymbol{k}\downarrow}\end{array})^{T}$, with $c_{\boldsymbol{k}\sigma=\uparrow,\downarrow}^{\dagger}$ ($c_{\boldsymbol{k}\sigma}$) the creation (annihilation) operator of a spin-$\sigma$ fermion with quasimomentum $\boldsymbol{k}=(k_x,k_y,k_z)$. $t_{\text{s},\zeta}$ is the hopping rate in the $\zeta$ spatial direction, $t_{\text{so}}$ is the SOC strength, $\sigma_{z}$ is the Pauli matrix, and $U$ is the interaction strength with the quantization volume given by $V$. Here the imaginary SOC may be implemented using spin-dependent dissipation~\cite{li2019observation}, non-reciprocal hopping~\cite{gou2020tunable}, or dissipative Raman processes~\cite{zhou2020dissipation}.

Hamiltonian (\ref{eq:origin_H}) is invariant under the combined transformation of the parity operator $\mathcal{P}: c_{\boldsymbol{k}\sigma}\rightarrow c_{-\boldsymbol{k}\sigma}$, and the time-reversal operator $\mathcal{T}: c_{\boldsymbol{k}\sigma}\rightarrow  [i\sigma_{y}]_{\sigma\sigma^{'}} c_{-\boldsymbol{k}\sigma^{'}} $ and $i\rightarrow -i $ (or equivalently, $i\mathcal{K}\sigma_{y}$ with the complex conjugation operator $\mathcal{K}$ and Pauli matrix $\sigma_{y}$ in the first quantization), but possesses neither $\mathcal{P}$ nor $\mathcal{T}$ symmetry separately. Notably, although the single-particle spectra of Hamiltonian (\ref{eq:origin_H}) are typically complex, the eigenenergies come in conjugate pairs, such that a non-interacting Fermi sea features a real Fermi energy.

\begin{figure}[tbp]
\includegraphics[width=8cm]{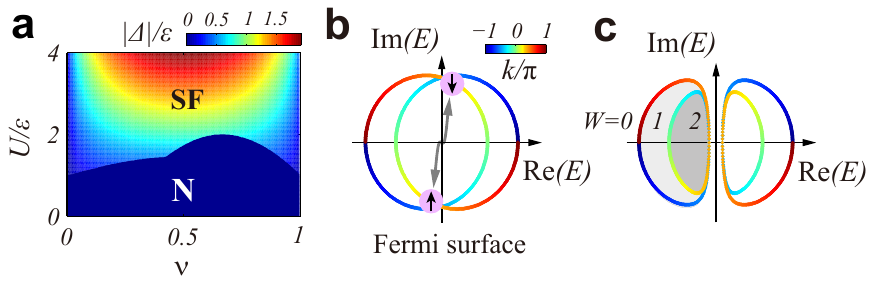}
\caption{{\bf Pairing in a non-Hermitian superfluid.} {\bf a} Phase diagram on the occupation fraction and interaction coefficient ($\nu$--$U$) plane, where the background colors indicate the order parameter $\Delta$. The self-consistent ground state of the system undergoes a first-order phase transition from the normal phase (N) to the superfluid phase (SF) when increasing $U$. {\bf b} and {\bf c}: Bogoliubov quasiparticle spectra in the limit $\Delta=0$ (the normal phase) {\bf b} and SF phase {\bf c} in the complex plane. As shown in {\bf b}, fermions in a potential Cooper pair are separated by a finite imaginary energy gap (vertical distance) on the Fermi surface, which underlies the first-order phase transition. The quasiparticle spectra are gapped in the SF phase [see {\bf c}], which implies the emergence of quantized spectral winding number $W$ for the occupied bands. {Different shades of grey in {\bf c} mark regions for different choice of reference energy. Here we take the hopping strength $t_{\text{s}}/\varepsilon=1$ and the SOC strength $t_{\text{so}}/\varepsilon=0.5$ with the unit of energy as the recoil energy $\varepsilon=\hbar^2 a^{-2}/(2m)$, where $\hbar$ is the Plank constant, $a$ is the lattice constant (unit of length) and $m$ is the mass of fermions.}}
\label{fig:BdG}
\end{figure}

{\bf Non-Hermitian BCS formalism.}
Pairing superfluidity has been investigated in open systems using non-Hermitian variations of the BCS formalism~\cite{kantian2009atomic, daley2009atomic, ghatak2018theory, kawabata2018parity, zhou2019enhanced, yamamoto2019theory}. In the spirit of these studies, we define the $s$-wave pairing order parameters $\Delta=-(U/V)\sum_{\boldsymbol{k}}\langle c_{-\boldsymbol{k}\downarrow}c_{\boldsymbol{k}\uparrow}\rangle$ and $\bar{\Delta}=-(U/V)\sum_{\boldsymbol{k}}\langle c_{\boldsymbol{k}\uparrow}^{\dagger}c_{-\boldsymbol{k}\downarrow}^{\dagger}\rangle$. Note that $\bar{\Delta}\neq \Delta^\ast$ for general non-Hermitian systems~\cite{yamamoto2019theory, ghatak2018theory}. The non-Hermitian BCS mean-field Hamiltonian is given by
\begin{equation}\label{eq:BCS}
\hat{H}_{\text{BCS}}=\sum_{\boldsymbol{k}}C_{\boldsymbol{k}}^{\dagger}h_{\boldsymbol{k}}C_{\boldsymbol{k}}+\sum_{\boldsymbol{k}}(\Delta c_{\boldsymbol{k}\uparrow}^{\dagger}c_{-\boldsymbol{k}\downarrow}^{\dagger}+\bar{\Delta}c_{-\boldsymbol{k}\downarrow}c_{\boldsymbol{k}\uparrow}),
\end{equation}
where a constant energy shift $V\Delta\bar{\Delta}/U$ is dropped.
The BCS Hamiltonian can be diagonalized as
$\hat{H}_{\text{BCS}}-\mu \hat{N}=\sum_{\boldsymbol{k}\sigma} E_{\boldsymbol{k}\sigma}\alpha_{\boldsymbol{k}\sigma}^{\dagger}\beta_{\boldsymbol{k}\sigma}$ (with the chemical potential $\mu$ and the total particle number operator $\hat{N}=\sum_{\boldsymbol{k}\sigma}c_{\boldsymbol{k}\sigma}^{\dagger}c_{\boldsymbol{k}\sigma}$)
through the Bogoliubov transformations
\begin{equation}\label{eq:quasi_particle}
\left(\begin{array}{c}
\beta_{\boldsymbol{k}\uparrow}\\
\beta_{\boldsymbol{k}\downarrow}
\end{array}\right)=U_{\boldsymbol{k}}\left(\begin{array}{c}
c_{\boldsymbol{k}\uparrow}\\
c_{-\boldsymbol{k}\downarrow}^{\dagger}
\end{array}\right),\left(\begin{array}{c}
\alpha_{\boldsymbol{k}\uparrow}^{\dagger}\\
\alpha_{\boldsymbol{k}\downarrow}^{\dagger}
\end{array}\right)=(U_{\boldsymbol{k}}^{-1})^{T}\left(\begin{array}{c}
c_{\boldsymbol{k}\uparrow}^{\dagger}\\
c_{-\boldsymbol{k}\downarrow}
\end{array}\right).
\end{equation}
Here $
U_{\boldsymbol{k}}=\left(\begin{array}{cc}
u_{\boldsymbol{k}} & \upsilon_{\boldsymbol{k}}\\
-\bar{\upsilon}_{\boldsymbol{k}} & u_{\boldsymbol{k}}
\end{array}\right)$,
with $u_{\boldsymbol{k}}=\sqrt{\frac{1}{2}(1+\frac{\xi_{\boldsymbol{k}}}{E_{\boldsymbol{k}}})}$, $\upsilon_{\boldsymbol{k}}=\sqrt{\frac{\Delta}{2\bar{\Delta}}(1-\frac{\xi_{\boldsymbol{k}}}{E_{\boldsymbol{k}}})}$, $\bar{\upsilon}_{\boldsymbol{k}}=\sqrt{\frac{\bar{\Delta}}{2\Delta}(1-\frac{\xi_{\boldsymbol{k}}}{E_{\boldsymbol{k}}})}$, $\xi_{\boldsymbol{k}}=-\sum_{\zeta=x,y,z}(t_{\text{s},\zeta}\cos k_{\zeta}+it_{\text{so}}\sin k_{x})-\mu$, and $E_{\boldsymbol{k}}=\sqrt{\bar{\Delta}\Delta+\xi_{\boldsymbol{k}}^{2}}$. Under the convention $\sqrt{z}\geq 0$ for $z\in \mathbb{C}$, $E_{\boldsymbol{k}\sigma}=\pm E_{\boldsymbol{k}}$ are respectively identified as the quasiparticle (positive) and quasihole (negative) dispersions, with the corresponding field operators satisfying $\{\alpha_{\boldsymbol{k}\sigma}^{\dagger},\beta_{\boldsymbol{k}^{'}\sigma^{'}}\}=\delta_{\sigma\sigma^{'}}\delta_{\boldsymbol{k}\boldsymbol{k}^{'}}$ and $\{\alpha_{\boldsymbol{k}\sigma},\beta_{\boldsymbol{k}^{'}\sigma^{'}}\}=0$.

The ground BCS state of the system is then constructed by filling the quasihole band $E_{\boldsymbol{k}\downarrow}$, and is captured by the density matrix $\rho_{\text{BCS}}=|\Psi_{\text{BCS}}\rangle\langle \tilde{\Psi}_{\text{BCS}}|$, with $|\Psi_{\text{BCS}}\rangle\propto\prod_{\boldsymbol{k}}(u_{\boldsymbol{k}}-\upsilon_{\boldsymbol{k}} c_{\boldsymbol{k}\uparrow}^{\dagger}c_{-\boldsymbol{k}\downarrow}^{\dagger})|\text{vac}\rangle$ and $|\tilde{\Psi}_{\text{BCS}}\rangle\propto\prod_{\boldsymbol{k}}(u_{\boldsymbol{k}}^{\ast} -\bar{\upsilon}_{\boldsymbol{k}}^{\ast}c_{\boldsymbol{k}\uparrow}^{\dagger}c_{-\boldsymbol{k}\downarrow}^{\dagger})|\text{vac}\rangle$. Such a treatment is equivalent to the zero-temperature limit of the Gibbs-state assumption $\rho_{G}=\exp[-\beta (\hat{H}_{\text{BCS}}-\mu \hat{N})]$ in Ref.~\cite{yamamoto2019theory}.

By substituting the BCS density matrix $\rho_{\text{BCS}}$ into the definitions of $\Delta=-(U/V)\sum_{\boldsymbol{k}}\text{Tr}(\rho_{\text{BCS}} c_{-\boldsymbol{k}\downarrow}c_{\boldsymbol{k}\uparrow})$ and the total particle number $N_{a}=\text{Tr}(\rho_{\text{BCS}}\hat{N})$, the self-consistent gap (top row) and number (lower row) equations are
\begin{equation}\label{eq:gapparticle_equation}
\left(\begin{array}{c}
1\\
\nu
\end{array}\right)=\frac{1}{2V}\sum_{\boldsymbol{k}}\left(\begin{array}{c}
U/E_{\boldsymbol{k}}\\
1-\xi_{\boldsymbol{k}}/E_{\boldsymbol{k}}
\end{array}\right),
\end{equation}
where the density $\nu=N_{a}/(2 V)$. Using $\xi_{\boldsymbol{k}}=\xi^\ast_{-\boldsymbol{k}}$ under the $\mathcal{PT}$ symmetry, an inspection of the summation in Eq.~(\ref{eq:gapparticle_equation}) reveals that the product $\Delta\bar{\Delta}$ must be real for this equation to hold.  Without loss of generality, we denote $\Delta=|\Delta|\text{e}^{i\theta}$ and $\bar{\Delta}=|\bar{\Delta}|\text{e}^{-i(\theta+n\uppi)}$, where $\theta$ is an arbitrary phase and $n\in \mathbb{Z}$. Interestingly,
under the $U(1)$ gauge transformation $c_{\boldsymbol{k}\sigma}\rightarrow \text{e}^{i\theta/2} c_{\boldsymbol{k}\sigma}$,
$\Delta$ and $\bar{\Delta}$ both become real numbers, which leads to $\mathcal{PT}|\Psi_{\text{BCS}}\rangle=|\Psi_{\text{BCS}}\rangle$, $\mathcal{PT}|\tilde{\Psi}_{\text{BCS}}\rangle=|\tilde{\Psi}_{\text{BCS}}\rangle$, and $\mathcal{PT}\rho_{\text{BCS}}(\mathcal{PT})^{-1}=\rho_{\text{BCS}}$. The BCS ground state thus preserves the $\mathcal{PT}$ symmetry up to a U(1) gauge transformation, which suggests that the BCS ground-state energy is necessarily real.

Furthermore, we find that the BCS ground state always lies within the sector $\Delta\bar{\Delta}>0$. Considering the fact that the gap and number equations are only dependent on the product $\Delta$ and $\bar{\Delta}$, rather than their relative ratio, we take $\Delta^{\ast}=\bar{\Delta}$, which does not affect the physical conclusions of our work.
For numerical calculations, we focus on a quasi-one-dimensional configuration, where the Fermi gas is tightly confined in the spatial directions perpendicular to that of the SOC, such that $t_{s,y,z}\ll t_{s,x}$.
Integrating out the transverse degrees of freedom and replacing $V^{-1}\sum_{\boldsymbol{k}}$ with $L_{x}^{-1}\sum_{k_{x}}$ ($L_x$ being the lattice size long the $x$ direction) in Eq.~(\ref{eq:gapparticle_equation}), we self-consistently solve the gap and number equations for $\{\Delta,\mu\}$, from which the BCS ground state as well as Bogoliubov quasiparticle spectra are constructed. For convenience, we drop the label $x$ in the following discussions. {Note that the residue degrees of freedom in the transverse direction differentiate our quasi-one-dimensional Fermi gas with an exact one-dimensional system where strong quantum fluctuations would render the mean-field treatment unreliable.}

{\bf Point-gap topological superfluid.}
In Fig.~\ref{fig:BdG}{\bf a}, we show the numerically calculated ground-state phase diagram (see additional discussions in the Supplementary Note 2). Unlike the conventional Hermitian case, where the superfluid phase transition is continuous, our model possesses a first-order phase boundary between the superfluid (SF) and the normal (N) phase, as evidenced by the plotted discontinuous color changes across the phase boundary.   Such a behavior originates from the competition between the pairing interaction and an imaginary gap introduced by the non-Hermitian SOC [see Fig.~\ref{fig:BdG}{\bf b} and \ref{fig:BdG}{\bf c}]. Note that the phase transition becomes continuous in the vacuum limit with the particle density $\nu=N_{a}/L\rightarrow 0$.

\begin{figure}[tbp]
\includegraphics[width=7.5cm]{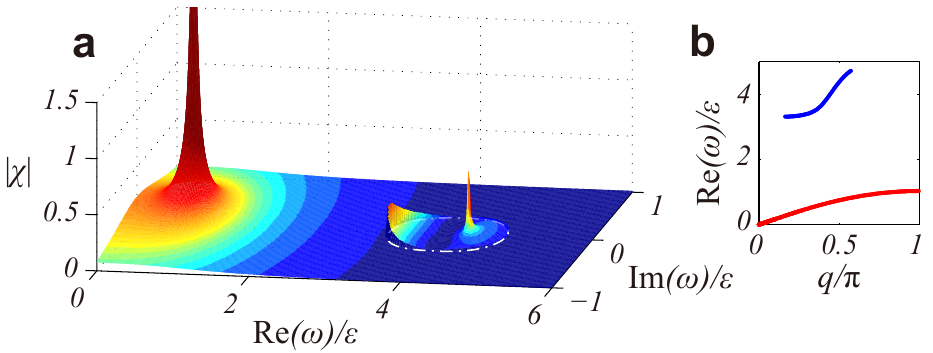}
\caption{{\bf Response function and collective modes.} {\bf a} Typical distribution of response function $\chi(q,\omega)$ on the complex frequency plane with a fixed perturbation momentum $q/\uppi=0.5$. The sharp peaks reflect the presence of collective modes. The surface colors are only employed to increase visibility. {\bf b} Dispersions of collective modes, obtained from the divergent sharp peaks plotted in panel {\bf a}. Note that the imaginary parts of the
spectra are zero here, as we set the spin-orbit coupling (SOC) strength $t_{\text{so}}$ smaller than the hopping coefficient $t_{\text{s}}$. The red lower (blue higher) branch locates outside (inside) the spectral loop $\omega=E_{k+q/2}+E_{k-q/2}$ with quasi-energy $E_{k}$, which is marked by the dash-dotted curve in {\bf a}. Here we take the parameters: hopping coefficient $t_{\text{s}}/\varepsilon=1$, SOC strength $t_{\text{so}}/\varepsilon=0.5$, occupation fraction $\nu=1/4$ and interaction coefficient $U/\varepsilon=4$ with the recoil energy $\varepsilon$, in our calculations. }
\label{fig:chi}
\end{figure}

In the superfluid phase, quasiparticle dispersions $E_{\boldsymbol{k}\sigma}$ form closed spectral loops on the complex plane, reminiscent of the eigenspectral point-gap topology  associated with the seminal non-Hermitian skin effects~\cite{gong2018topological,kawabata2019symmetry,bergholtz2019exceptional,okuma2020topological,zhang2020correspondence,pan2020point}.  The spectral winding number characterizing the ground-state point-gap topology is given by
\begin{equation}\label{eq:WN}
W(\Omega)=\frac{1}{2\uppi \text{i}}\int dk\frac{\partial}{\partial k}\arg[E_{k\downarrow}(k)-\Omega],
\end{equation}
where $\Omega$ is the reference energy. In contrast to the normal phase, where the excitation spectra are gapless and $W$ is ill-defined, $W$ takes quantized values in the superfluid phase. As illustrated in Fig.~\ref{fig:BdG}{\bf c}, $W(\Omega)$ can take quantized values of $0$, $1$, or $2$,
when $\Omega$ is chosen within different regimes.
The absolute value of $W$ indicates the degeneracy of edge modes with eigenenergy $\Omega$, under a semi-infinite boundary condition ~\cite{okuma2020topological,zhang2020correspondence}(also see the Supplementary Note 1). This implies that the BCS state possesses not only pairing order parameter but also nontrivial point-gap topology, and thus represents a point-gap topological superfluid state.
Whereas it is naturally expected that quasiparticle excitations of the superfluid would similarly be localized at the boundaries under an open boundary condition~\cite{yao2018edge,kunst2018biorthogonal,lee2016anomalous,xiong2018does,alvarez2018topological,macdonald2018phase,lee2019anatomy,li2019geometric,yokomizo2019non,kawabata2019symmetry,
hofmann2020reciprocal,xiao2020non,helbig2020generalized,bergholtz2019exceptional}, we instead focus here on the physics of collective modes, where the point-gap topological nature of the superfluid has a dramatic impact.

{\bf Spontaneous $\mathcal{PT}$-symmetry breaking of AB modes.}
The spontaneous breaking of U(1) gauge symmetry by the pairing order generally leads to the emergence of gapless AB collective modes, which manifest themselves as the divergence in the linear response.
We extend the conventional dynamic BCS theory~\cite{combescot2006collective,combescot1982dispersion} into the non-Hermitian regime, and derive the density response function
\begin{equation}\label{eq:response}
\chi(q,\omega)=\frac{1}{4\uppi}(I^{''}-\left|\Delta\right|^{2}\frac{I^{2}I_{11}+\omega^{2}I^{'2}I_{22}-2\omega^{2}I_{12}I^{'}I}{I_{11}I_{22}-\omega^{2}I_{12}^{2}}),
\end{equation}
where  the response function $\chi(q,\omega)$ characterizes the density fluctuation of the superfluid to a small external perturbation of frequency $\omega$ and momentum $q$, with the definitions of integrals $\{I, I^{'},I^{''},I_{11},I_{12},I_{22}\}$ as functions of $\omega$ and $q$ {\color{black}(see the details in Methods and the Supplementary Note 3)}.
To unravel the complete response feature of our model, we extend the definition of $\chi$ into the complex-frequency regime~\cite{xue2020non,li2021quantized}, which corresponds to the linear response of damped/amplified perturbations, when the frequency deviates from the real axis.
Similar to the Hermitian case, the first term on the right hand side of Eq.~(\ref{eq:response}) represents the linear-response results from the standard BCS theory, and the second term, being proportional to $|\Delta|^2$, represents contributions from quantum fluctuations of the pairing field that are responsible for the AB collective modes.

{
In general, $\chi$ has two types of poles: poles of $I^{''}$ which arise from the breaking of Cooper pairs into Bogoliubov quasiparticles; and poles of the second term in Eq.~(\ref{eq:response}), which satisfy $I_{11}I_{22}=\omega^{2}I_{12}^{2}$, and originate from the AB collective modes~\cite{combescot2006collective}. In fact, poles of the first type generally exist in the integrands of $\{I, I^{'}, I^{''}, I_{11},I_{12},I_{22}\}$ which appear to be singular at $\omega=E_{k+q/2}+E_{k-q/2}$ (see the Supplementary Note 3). In a Hermitian BCS state,
most of these poles are removable singularities of the integral, with the only irremovable ones located at the extremal frequencies, i.e., when $\omega=\text{max}(E_{k+q/2}+E_{k-q/2})$ or $\omega=\text{min}(E_{k+q/2}+E_{k-q/2})$ with $k\in [-\uppi,\uppi)$. These irremovable singularities mark the thresholds for the pair-breaking process.
By contrast, in our non-Hermitian BCS state, since the frequency $\omega$ is extended to the complex regime, the spectral winding of quasiparticles in general guarantees that the extremal conditions of the real and complex components of $\omega=E_{k+q/2}+E_{k-q/2}$ should not be satisfied simultaneously. This implies that the extremal conditions can never be met, and the singularities are all removable. For example, the integrals around the singular points with minimal real frequencies can be approximately rewritten as $\int_{-\delta}^{\delta} dk (k^2+\text{i}ck)^{-1}=(\text{i}c)^{-1} \int_{-\delta}^{\delta} dk [k^{-1}-(k+ic)^{-1}]\approx 2\delta/c^{2}$, which implies that the integrals are not divergent. It follows that poles of the first type completely disappear in our non-Hermitian setting.}

\begin{figure}[tbp]
\includegraphics[width=8.8cm]{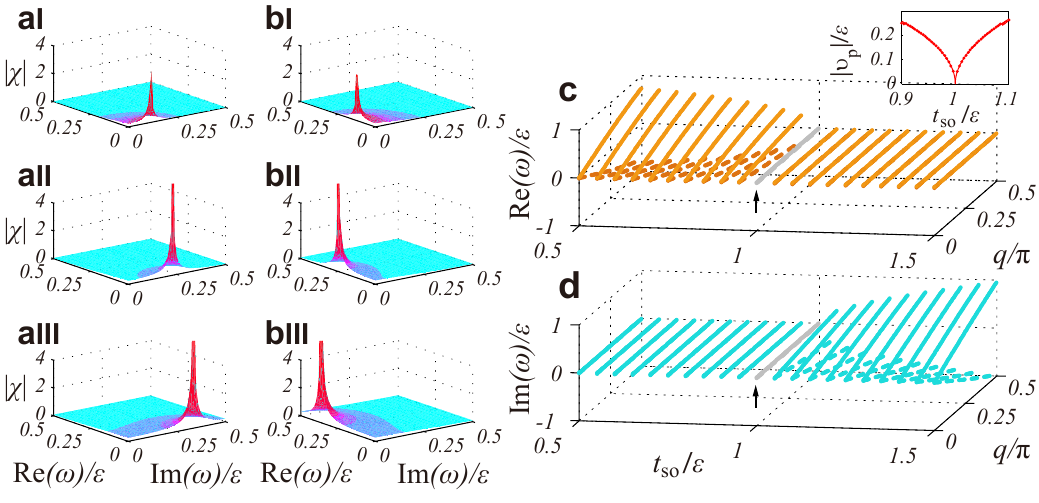}
\caption{{\bf PT transition in the collective modes.} Response function $\chi(q,\omega)$ in the low-frequency regime for spin-orbit coupling (SOC) strength $t_{\text{so}}/\varepsilon=0.8$ ($t_{\text{so}}/\varepsilon=1.2$) in the column {\bf aj} ({\bf bj}), with $j=I,II,III$. Here {\bf aI} and {\bf bI} correspond to the perturbation momentum $q/\uppi=0.1$; {\bf aII} and {\bf bII}: $q/\uppi=0.2$; {\bf aIII} and {\bf bIII}: $q/\uppi=0.3$. {{\bf c} and {\bf d}: Spectra of AB mode $\omega_{AB}$ (solid) and its hole partner $-\omega_{AB}$ (dashed) coalesce at the critical point (indicated by arrow), which corresponds to the real-to-complex transition point illustrated in panels {\bf a}s and {\bf b}s.
Locations of the critical point in {\bf c} and {\bf d} are indicated by gray lines, which are not poles of $\chi(q,\omega)$ and do not correspond to the spectra of collective modes.
Inset of {\bf c}:} numerically calculated speed of sound (blue dots) near the critical point, which is well-fitted by a power-law function $\propto|t_{\text{s}}-t_{\text{so}}|^{\gamma}$ (red curves), with a critical exponent $\gamma\approx 1/2$, where $t_{\text{s}}$ is the hopping coefficient. Here we take $t_{\text{s}}/\varepsilon=1$, interaction strength $U/\varepsilon=4$ and occupation fraction $\nu=1/4$, with the recoil energy $\varepsilon$. }
\label{fig:collective_mode_scan_t2}
\end{figure}

In Fig.~\ref{fig:chi}{\bf a}, we show the typical landscape of $\chi$ on the complex frequency plane. Two separate poles (both of the second type) can be identified, {one in the low-frequency regime, located on either the real of imaginary axis; the other in the high-frequency regime, lying within the spectral loop $\omega=E_{k+q/2}+E_{k-q/2}$. Note that these are contrasted with the poles of $I^{''}$, which only appear on the spectral loop $\omega=E_{k+q/2}+E_{k-q/2}$, not within.}
Both poles are characterized by $I_{11}I_{22}=\omega^{2}I_{12}^{2}$, and therefore contribute toward the AB collective modes.
{ Given the location of the high-frequency pole, the high-frequency AB modes are then always gapped.}
To understand the gapless phonon excitations, we thus focus on the low-frequency regime [see the lower branch in Fig.~\ref{fig:chi}{\bf b}].

In Fig.~\ref{fig:collective_mode_scan_t2}, we show the response function in the low-frequency regime, with either small ({\bf aI}, {\bf aII} and {\bf aIII}) or large ({\bf bI}, {\bf bII} and {\bf bIII}) SOC strength. The location of sharp peaks satisfy $I_{11}I_{22}=\omega^{2}I_{12}^{2}$ (at the obvious pole of the response function), from which we solve for the
dispersion of AB modes $\omega_{AB}(q)$ for different SOC strengths [see Fig.~\ref{fig:collective_mode_scan_t2}{\bf c} and \ref{fig:collective_mode_scan_t2}{\bf d}]. Note that the resultant dispersions of AB modes change from purely real for $t_{\text{so}}<t_{\text{s}}$, to purely imaginary for $t_{\text{so}}>t_{\text{s}}$, thus indicating an intriguing transition point.  This is in contrast to the BCS ground state that always features a real energy. Note that the spectra of the higher branch of the collective modes [see Fig.~\ref{fig:chi}{\bf b}] obtained from the sharp peaks plotted in Fig.~\ref{fig:chi}{\bf a}],  are also real, regardless of the value of $t_{\text{so}}$. Thus, an emergent $\mathcal{PT}$-symmetry breaking occurs in the lower branch of the collective modes, at the critical point $t_{\text{so}}=t_{\text{s}}$. Close to the critical point, the speed of sound $\upsilon_{p}=\partial \omega_{AB}/\partial q\big|_{q=0}$, which characterizes the speed of propagation for phonon modes, rapidly vanishes toward a non-analytic kink at the transition point [see inset of Fig.~\ref{fig:collective_mode_scan_t2}{\bf c}], confirming the existence of a quantum phase transition~\cite{sachdev2007quantum}.
{ Furthermore, under a low-frequency, small-momentum expansion, the equation $I_{11}I_{22}=\omega^{2}I_{12}^{2}$ is reduced to $\omega=\pm \upsilon_{p}$, with $\upsilon^2_{p}>0$ ($\upsilon^2_{p}<0$) in the $\mathcal{PT}$ unbroken (broken) phase (see the Methods for more information). Numerical calculations then reveal that, at the critical point, $\upsilon_{p}$ vanishes with a critical exponent $1/2$, consistent with a direct numerical fit in Fig.~\ref{fig:response_EP}.
}

The softening of phonon mode derives from the coalescence of the AB mode and its hole partner [see Fig.~\ref{fig:collective_mode_scan_t2}{\bf c} and \ref{fig:collective_mode_scan_t2}{\bf d}]. In contrast to the mere merging of eigenmodes in single-particle $\mathcal{PT}$-symmetric systems, the particle and hole modes annihilate each other at the critical point. This is also in contrast to previous studies on the coalescence of the Bogoliubov modes at the critical exceptional point of a two-component Bose-Einstein condensate~\cite{hanai2019non,hanai2020critical,fruchart2020phase}, where phonon modes with linear dispersion persist at the critical point. Indeed, a direct consequence of the annihilation of AB modes here
is the complete absence of response to low-frequency perturbations, which, as we show below, is associated with the spectral point-gap topology.

{\bf Critical phase.}
Here we focus on system's behavior precisely at the critical point $t_{\text{so}}=t_{\text{s}}$.  The real-to-complex transition at this point indicates that the low-frequency branch of the response function vanishes there, such that collective modes of the system are entirely determined by the high-frequency branch [see Fig.~\ref{fig:response_EP}{\bf a}]. We then have access to a peculiar scenario.  As illustrated in the inset of Fig.~\ref{fig:response_EP}{\bf a}, the total response function $\chi$ completely vanishes outside the spectral loop $\omega=E_{k+q/2}+E_{k-q/2}$, whose shapes are shown in Fig.~\ref{fig:response_EP}{\bf b}, even though contributions from the BCS theory (blue) and order-parameter fluctuations (black) remain finite (see the shaded region). Physically, such a behavior suggests a critical phase that is immune to low-frequency perturbations. Remarkably, the total absence of linear response at the critical point can be analytically proven by changing the integrals in Eq.~(\ref{eq:response}) into contour integrals on the complex plane, with the transformation $z=\text{e}^{\text{i}k}$. Given the spectral-loop structures of the Bogoliubov quasiparticles, all the integrals can be performed analytically through the Cauchy's theorem (see the Supplementary Note 4 for the calculations). As such, the robustness of the critical phase is linked to the spectral point-gap topology of the Fermi quasiparticles.

\begin{figure}[tbp]
\includegraphics[width=7.3cm]{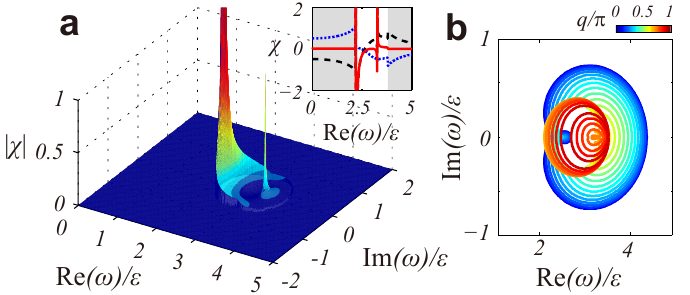}
\caption{{\bf Critical response.} {\bf a} Typical distribution of response function $|\chi|$ on the complex frequency plane at the critical point (take perturbation momentum $q/\uppi=0.5$ for example). Inset: the sectional view of $\chi$ along the real axis, where the red solid, blue dotted and black dashed curves denote the total, the part arising from the simple Bardeen-Cooper-Schrieffer (BCS) theory, and the part arising from the order-parameter fluctuations of $\chi$, respectively. { We see that $\chi$ vanishes outside the gapped spectral loops of $\omega=E_{k+q/2}+E_{k-q/2}$ with quasi-energy $E_{k}$ (see also the shaded regions in the inset of {\bf a}), whose shapes for different $q$ are shown in {\bf b}. As a result, $\chi$ always vanishes for $\omega$ outside all such spectral loops for any $q$.} Here we set the interaction strength $U/\varepsilon=4$, hopping coefficient $t_{\text{s}}/\varepsilon=t_{\text{so}}/\varepsilon=1$ and occupation coefficient $\nu=1/4$, where $t_{\text{so}}$ is the spin-orbit-coupling (SOC) strength. }
\label{fig:response_EP}
\end{figure}

Physically, the disappearance of linear response at the critical point can be understood through the exotic behavior of the BCS theory at the critical point. Explicitly, at the critical point, the quasiparticle spectrum is given by $E(z)=\sqrt{|\Delta|^{2}+(t_{\text{s}} z+\mu)^{2}}$, with $z=\text{e}^{\text{i}k}$ and $k\in[0,2\uppi)$, which is analytic on the complex plane of $z$. The gap and number equations are therefore only determined by the residue near $z=0$, and can be reduced to
\begin{equation}\label{eq:BCS_independent_dispersion}
\frac{1}{U}=\frac{1}{2\sqrt{|\Delta|^{2}+\mu^{2}}}, \quad\nu=\frac{1}{2}(1+\frac{\mu}{\sqrt{|\Delta|^{2}+\mu^{2}}}).
\end{equation}
The above equations can be analytically solved, with the solutions $\mu=U(\nu-\frac{1}{2})$ and $|\Delta|=U\sqrt{\nu(1-\nu)}$. More importantly, since the BCS theory at the critical point is dispersionless, for perturbations with frequencies outside the spectral loop $\omega=E_{k+q/2}+E_{k-q/2}$ in the complex plane, their impact on the system would be the same as that of a zero-momentum perturbation, which cannot lead to any fluctuations. The critical phase is only responsive to perturbations with frequencies lying within the spectral loop $\omega=E_{k+q/2}+E_{k-q/2}$.

{The complete suppression of the response function here is to be contrasted with the {partial} suppression in gapped systems. Therein, the response to an external perturbation at a frequency inside the gap is generally suppressed but does not completely vanish. Indeed, the coupling with bulk modes contributes perturbatively to the response function which, to leading order, becomes
inversely proportional to the detuning of the perturbation frequency with respect to the bulk spectra.
A vanishing response function outside the spectral loop is therefore entirely nontrivial.}

\begin{center}{\bf Conclusion}\end{center}

We have uncovered an emergent $\mathcal{PT}$-symmetry breaking in the collective AB modes of a Fermi superfluid, and characterized in detail the exotic many-body critical phenomena at the transition point. {Despite the complex quasiparticle spectra, the superfluid state is generally stabilized by the finite pairing gap. However, in the $\mathcal{PT}$-broken regime where the low-energy AB modes become purely imaginary, the superfluid may become dynamically unstable, and give way to more exotic dynamic phenomena.}
In previous studies, $\mathcal{PT}$-symmetry breaking in the superconductivity fluctuations has been reported in superconducting wires~\cite{rubinstein2007bifurcation,chtchelkatchev2012stimulation}. Their starting point, however, is the phenomenological Ginsburg-Landau field theory, and the dominant fluctuations therein originate from Cooper-pair breaking, rather than the AB modes discussed here.
The topologically robust critical phase discussed in this work opens up a new avenue toward the preparation of perturbation-free quantum states under $\mathcal{PT}$ symmetry.
Further, while we focus on a quasi-one-dimensional configuration, emergent $\mathcal{PT}$-symmetry breaking should also occur in higher dimensions, where the impact of dimensionality on the many-body critical phenomena would be an interesting open question for future studies.

\begin{center}{\bf Methods}\end{center}

{\color{black}
{\bf Linear response function under the dynamic BCS theory.}
The linear response function $\chi(q,\omega)$, which characterizes the dynamics of superfluid state in the presence of small external perturbations, is derived under the dynamic BCS theory~\cite{combescot2006collective,combescot1982dispersion}. In the Hermitian limit, the dynamic BCS theory yields the same response function with that from the diagrammatical approach, but the former has a more straightforward physical picture~\cite{combescot2006collective}.

Our starting point is the non-Hermitian BCS Hamiltonian
\begin{equation}\label{eq:H_BCS}
\begin{split}
\hat{H}_{0}=&\hat{H}_{\text{BCS}}-\mu \hat{N}=\sum_{k}(\begin{array}{cc}
c_{k\uparrow}^{\dagger} & c_{-k\downarrow}\end{array})\left(\begin{array}{cc}
\xi_{k} & \Delta\\
\bar{\Delta} & -\xi_{k}
\end{array}\right)\left(\begin{array}{c}
c_{k\uparrow}\\
c_{-k\downarrow}^{\dagger}
\end{array}\right)\\
&=\sum_{k}(\begin{array}{cc}
c_{k1}^{\dagger} & c_{k2}^{\dagger}\end{array})\hat{\epsilon}_{k}^{0,T}\left(\begin{array}{c}
c_{k1}\\
c_{k2}
\end{array}\right),
\end{split}
\end{equation}
where $c_{k1}=c_{k\uparrow}$, $c_{k2}=c_{-k\downarrow}^{\dagger}$, and $ \hat{\epsilon}_{k}^{0}=\left(\begin{array}{cc}
\xi_{k} & \bar{\Delta}\\
\Delta & -\xi_{k}
\end{array}\right)$.
To show that the response function is only dependent on $\bar{\Delta}\Delta$, as we claimed in the main text (which is not apparent here), we regard $\Delta^{\ast}$ and $\bar{\Delta}$ as different quantities throughout the derivation here.
Assuming the density fluctuations are coupled with external perturbations of frequency $\omega$ and momentum $q$, the perturbation-fluctuation Hamiltonian is given by
\begin{equation}\label{eq:perturbation_fluctuation}
\delta H=\sum_{k\lambda\lambda^{'}}c_{k+q/2\lambda}^{\dagger}(\delta\hat{\epsilon}^{T})_{\lambda\lambda^{'}}c_{k-q/2\lambda^{'}}+h.c.,
\end{equation}
with $\delta\hat{\epsilon}=\left(\begin{array}{cc}
\delta F & \delta\bar{\Delta}\\
\delta\Delta & -\delta F
\end{array}\right)$, where $\delta F$ and $\delta\Delta$ respectively denote the external perturbations and the fluctuations of pairing field.
Then the dynamic BCS Hamiltonian is given by
\begin{equation}
 \hat{H}_{tot}=\hat{H}_{0}+\delta\hat{H},
\end{equation}
which also satisfies the parity symmetry $\mathcal{P}\hat{H}_{tot}\mathcal{P}^{-1}=\hat{H}_{tot}^{\dagger}$.  The density response function is defined as
\begin{equation}
\chi\left(q,\omega\right)=\delta n/\delta F,\quad\delta n=L^{-1}\sum_{k}\left(\delta n_{k}\right)_{11},
\end{equation}
where $(\delta\hat{n}_{k})_{\lambda\lambda^{'}=1,2}=\langle c_{k-q/2,\lambda}^{\dagger} c_{k+q/2,\lambda^{'}}\rangle$.

Considering the time evolution of $|\Psi_{\text{BCS}}\rangle$ under $\hat{H}_{\text{tot}}$ in the Schr\"odinger picture, along with $\rho_{\text{BCS}}=|\Psi_{\text{BCS}}\rangle\langle\Psi_{\text{BCS}}|\mathcal{P}$ ($\mathcal{P}$ being the parity operator), we write down the Heisenberg equation $i\partial\langle \delta \hat{n}_{k}\rangle /\partial t=\langle\delta \hat{n}_{k}\hat{H}_{t}-\mathcal{P}^{-1}\hat{H}_{t}^{\dagger}\mathcal{P}\delta \hat{n}_{k}\rangle$, which can be simplified as
\begin{equation}\label{eq:kinetic_equation}
\omega\delta\hat{n}_{k}\approx\delta\hat{n}_{k}\hat{\epsilon}_{k+q/2}^{0}-\hat{\epsilon}_{k-q/2}^{0}\delta\hat{n}_{k}+\hat{n}_{k-q/2}^{0}\delta\hat{\epsilon}_{k}-\delta\hat{\epsilon}_{k}\hat{n}_{k+q/2}^{0}.
\end{equation}
Here we follow the spirit of linear response and write $\delta\hat{n}_{k}(t)\sim e^{i\omega t}\delta\hat{n}_{k}(0)$, where the matrix elements of the density operator $\hat{n}_m^0$ is given by $(\hat{n}_{k}^{0})_{\lambda\lambda^{'}}=\langle c_{k\lambda}^{\dagger} c_{k\lambda{'}}\rangle$.

{Equation (\ref{eq:kinetic_equation}) is the kinetic equation that characterizes the fluctuation dynamics, formally the same as that in the Hermitian case~\cite{combescot2006collective}. Such a formal invariance is due to the presence of an $\eta$-pseudo-Hermiticity $\hat{H}^{\dagger}_{\text{tot}}=\eta \hat{H}_{\text{tot}}\eta^{-1}$ with $\eta=\mathcal{P}$ in our model. Hence, by writing the density matrix as $\rho_{\text{BCS}}=|\Psi_{\text{BCS}}\rangle\langle\Psi_{\text{BCS}}|\mathcal{P}$, the expectation value of an operator $\hat{A}$ is given by $\langle\Psi_{\text{BCS}}|\eta \hat{A}|\Psi_{\text{BCS}}\rangle$. The formal invariance of the dynamic equation is therefore guaranteed by that of the pseudo-Hermitian quantum mechanics~\cite{mostafazadeh2002pseudoI,mostafazadeh2002pseudoII,mostafazadeh2002pseudoIII,mostafazadeh2002pseudo,gardas2016non}.}

To unravel the complete response feature of our model, we discuss the response dynamics in the whole the complex-frequency regime~\cite{xue2020non,li2021quantized}, which corresponds to the linear response of damped/amplified perturbations when the frequency deviates from the real axis. The kinetic equation can be consistently solved with the dynamic extension of gap equation,
\begin{equation}\label{eq:dynamic_qp}
\sum_{k}\left[\left(\delta\hat{n}_{k}\right)_{12}+\frac{\delta\bar{\Delta}}{2E_{k}}\right]=0,\quad\sum_{k}\left[\left(\delta\hat{n}_{k}\right)_{21}+\frac{\delta\Delta}{2E_{k}}\right]=0,
\end{equation}
which gives the expression of linear response function $\chi(q,\omega)=\delta \hat{n}_{11}/\delta \hat{\epsilon}_{11}$. As shown in Supplementary Note 3, the derivation process of $\chi(q,\omega)$ is lengthy but is straightforward.
}

{\bf Low-frequency expansion of the collective modes.} The dispersion of collective modes is determined by the equation $I_{11}I_{22}=\omega^{2}I_{12}^{2}$, and can be analytically expanded in the low-frequency, small-momentum limit following Ref.~\cite{combescot2006collective}. We start by rewriting the integral $I_{11}$ as
\begin{equation}
I_{11}=\frac{1}{2}\int dk\frac{E_{+}+E_{-}}{E_{+}E_{-}}\frac{\omega^{2}-(\xi_{+}-\xi_{-})^{2}}{(E_{+}+E_{-})^{2}-\omega^{2}}.
\end{equation}
In the low-frequency, small-momentum limit, $\xi_{\pm}\approx\xi_{k}\pm\eta_{k} q$, with $\eta_{k}=t_{s}\sin(k)-it_{so}\cos(k)$. Therefore,
\begin{equation}
I_{11}\approx \frac{J_{a}}{4}\omega^{2}-J_{b}q^{2},
\end{equation}
where $J_{a}=\int dk \,1/E_{k}^3$ and $J_{b}=\int dk \,\eta_{k}^2/E_{k}^3$. Similarly, we also have $I_{22}\approx I_{11}-J_{a}|\Delta|^2$, and $I_{12}=J_{c}/2$ with $J_{c}=\int dk \xi_{k}/E_{k}^3$. The equation $I_{11}I_{22}=\omega^{2}I_{12}^{2}$ then leads to
\begin{equation}
\omega=\upsilon_{p} q,\quad \upsilon_{p}=\pm\sqrt{\frac{J_{a}J_{b}|\Delta|^2}{J_{c}^2+J_{a}^2|\Delta|^2}}.\label{eq:Svp}
\end{equation}
In Fig.~\ref{fig:SM_phonon_velocity}{\bf a}, we plot the numerically calculated expansion coefficients $J_{a,b,c}$, which are all real numbers since
$\xi_{-k}=\xi_{k}^{\ast}$ and $\eta_{-k}=-\eta_{k}^{\ast}$.

\begin{figure}[tbp]
\includegraphics[width=8.5cm]{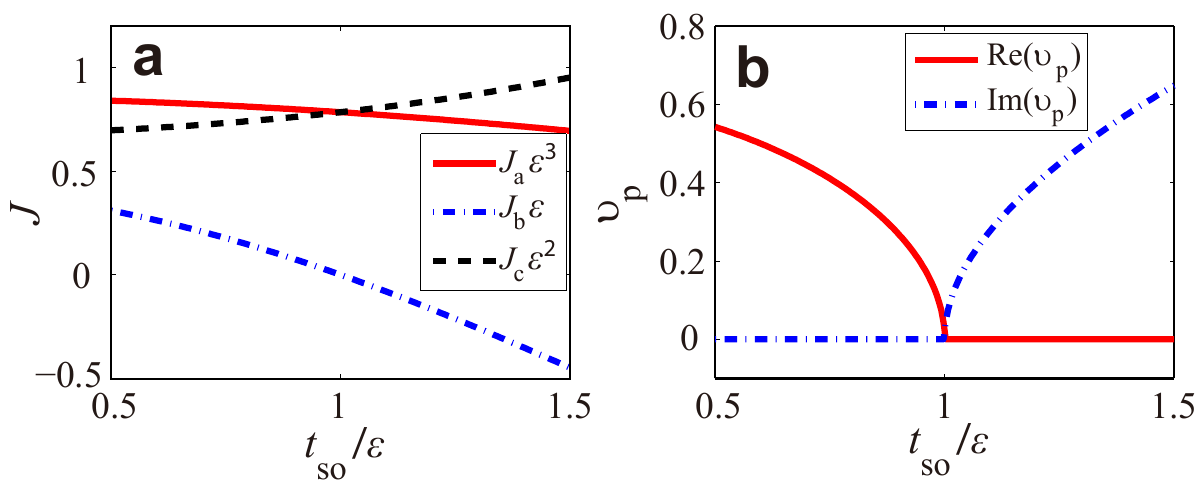}
\caption{{{\bf Softening of phonon velocity at the critical point.} Numerically evaluated {\bf a} expansion coefficients $J_{a,b,c}$ and {\bf b} phonon velocity $\upsilon_{p}$ as functions of the spin-orbit coupling (SOC) strength $t_{\text{so}}$. Other parameters are the same as those in Fig.~\ref{fig:response_EP}. Here we fix the interaction strength $U/\varepsilon=4$, hopping coefficient $t_{\text{s}}/\varepsilon=1$ and occupation coefficient $\nu=1/4$.}}
\label{fig:SM_phonon_velocity}
\end{figure}

{
We now study the collective spectra at the critical point, with $t_{so}=t_{s}$, $\xi_{k}=-t_{s}z-\mu$, and $\eta_{k}=-it_{s}z$, where $z:=e^{\text{i}k}$. From the residue theorem, we have
\begin{equation}
\begin{split}
J_{a}=&-\text{i}\int_{|z|=1} dz \frac{1}{z[\sqrt{|\Delta|^2+(t_{s}z+\mu)^2}]^3}=\frac{2\uppi}{[\sqrt{|\Delta|^2+\mu^2}]^3},\\
J_{b}=&-\text{i}\int_{|z|=1} dz \frac{(-\text{i}t_{s}z)^2}{z[\sqrt{|\Delta|^2+(t_{s}z+\mu)^2}]^3}=0,\\
J_{c}=&-\text{i}\int_{|z|=1} dz \frac{-\mu-t_{s}z}{z[\sqrt{|\Delta|^2+(t_{s}z+\mu)^2}]^3}=\frac{-2\mu\uppi}{[\sqrt{|\Delta|^2+\mu^2}]^3}.
\end{split}
\end{equation}
Note that we have used the fact that $E_{k}$ is gapped above the BCS ground state. This means $J_{a}$ and $J_{c}$ are finite and positive at the critical point (note that $\mu<0$), while $J_{b}$ vanishes. Further, we calculate the derivative of $J_{b}$ at the critical point $t_{so}=t_{s}$
\begin{equation}
\begin{split}
\frac{\partial J_{b}}{\partial t_{so}}&=\int dk \frac{2\eta_{k}}{E_{k}^3}\frac{\partial \eta_{k}}{\partial t_{so}}+\int dk \frac{-3\eta_{k}}{E_{k}^4}\frac{\partial E_{k}}{\partial t_{so}}\\
&=-\text{i}\int_{|z|=1} dz \frac{2\eta(z)}{zE^3(z)}[-\frac{\text{i}}{2}(z+z^{-1})]\\
&+\text{i}\int_{|z|=1} dz \frac{3\eta^2(z)}{zE^{4}(z)}\frac{1}{E(z)}[|\Delta|\frac{\partial |\Delta|}{\partial t_{so}}\\
&-(\frac{\partial \mu}{\partial t_{so}}+\frac{z+z^{-1}}{2})\xi(z)]\\
&=\frac{-2\uppi t_{s}}{(\sqrt{\mu^2+|\Delta|^2})^{3}}<0.
\end{split}
\end{equation}
This implies $J_{b}$ changes its sign from positive to negative at the critical point, which, according to Eq.~(\ref{eq:Svp}), is responsible for the transition of $\upsilon_{p}$ from real to imaginary values at the critical point.
}

{
We show the numerically calculated $\upsilon_{p}$ in Fig.~\ref{fig:SM_phonon_velocity}{\bf b}. Clearly,
$\upsilon_{p}$ softens to zero at the critical point $t_{so}=t_{s}$, as discussed in the main text, and becomes imaginary in the $\mathcal{PT}$-broken regime. Since $J_{b}$ is linear in $|t-t_{so}|$ close to the critical point, $\nu_{p}$ indeed scales as $|t-t_{so}|^{\frac{1}{2}}$, consistent with numerical fits in the inset of Fig.~\ref{fig:collective_mode_scan_t2}{\bf c}. Note that we only show the positive branches of $\upsilon_{p}$ in Fig.~\ref{fig:SM_phonon_velocity}{\bf b}. The negative branches, being the hole partners of AB modes, are symmetric to the positive branches.}

\bibliographystyle{naturemag}
\normalem
% this command is used to eliminate the underline below journals
\bibliography{NHTS_ref_CP}

\vspace{0.3cm}
{\bf Acknowledgements}
The authors wish to thank Dr. Sen Mu and Dr. Ryo Hanai for very helpful discussions. J.-S. P. acknowledges the support from the National Natural Science Foundation of China (Grant No. 11904228) before he joined NUS. W. Y. acknowledges support by the National Natural Science Foundation of China (Grant No. 11974331), and the National Key R\&D Program (Grant Nos. 2016YFA0301700, 2017YFA0304100). J. G.  acknowledges funding support from Singapore National Research Foundation Grant No. NRF- NRFI2017-04 (WBS No. R-144-000-378-281).

\vspace{0.2cm} {\bf Author contributions}
J.-S. P. performed the theoretical calculations with input from W. Y. and J. G.. All of the authors designed the project, analyzed the results and wrote the paper. J.G. supervised this project.

\vspace{0.2cm} {\bf Additional information}
Correspondence and requests for materials should be addressed to Jiangbin Gong (phygj@nus.edu.sg), Wei Yi (wyiz@ustc.edu.cn) or Jian-Song Pan (panjsong@gmail.com).

\vspace{0.2cm} {\bf Competing interests}
The authors declare no competing interests.

\vspace{0.2cm} {\bf Data availability}
The data that support the plots within this paper are available from the corresponding
authors upon reasonable requests.

\vspace{0.2cm} {\bf Code availability}
The codes that support the plots within this paper are available from the corresponding
authors upon reasonable requests.

\end{document}